\documentclass[a4paper,11pt]{article}

\usepackage{pos}
\usepackage{subfig} 
\usepackage{wrapfig}
\usepackage{multicol}
\usepackage{lineno}

\let\OLDthebibliography\thebibliography
\renewcommand\thebibliography[1]{
  \OLDthebibliography{#1}
  \setlength{\parskip}{0pt}
  \setlength{\itemsep}{0pt plus 0.3ex}
}

\title{Time Variation in the TeV Cosmic Ray Anisotropy with IceCube and Energy Dependence of the Solar Dipole}

\ShortTitle{Time Variation in the TeV Cosmic Ray Anisotropy}

\author{The IceCube Collaboration \\{\normalsize \normalfont(a complete list of authors can be found at the end of the proceedings)}\\}

\emailAdd{pzilberman@wisc.edu}
\emailAdd{juancarlos@icecube.wisc.edu}
\emailAdd{paolo.desiati@icecube.wisc.edu}

\abstract{

There is an observed anisotropy in the arrival direction distribution of cosmic rays in the TeV-PeV regime with variations on the scale of one part in a thousand. While the origin of this anisotropy is an open question, a possible factor is cosmic-ray interactions with interstellar and heliospheric magnetic fields. These magnetic fields may change over time - for example, due to changes in solar activity throughout its 11-year solar cycle. The cosmic-ray anisotropy can reflect these time-dependent magnetic fields. In addition to these speculative sources, there are several known sources of time variation in this anisotropy, such as the Compton-Getting Effect from the Earth’s orbital motion. We discuss a preliminary study with limited statistics of time variation undertaken by the IceCube Neutrino Observatory, including a measurement of the Compton-Getting Effect as well as a general, model-independent search for other time variations. Further, we use the Compton-Getting Effect to present a preliminary measurement of the cosmic-ray spectral index as a function of energy below the knee.

\vspace{4mm}

{\bfseries Corresponding authors:}
Perri Zilberman$^{1*}$, 
Juan Carlos Díaz Vélez$^{1}$, 
Paolo Desiati$^{1}$\\
{$^{1}$ \itshape Dept. of Physics and Wisconsin IceCube Particle Astrophysics Center, University of Wisconsin{\textemdash}Madison, Madison, WI 53706, USA}\\
[4mm]
$^*$ Presenter
}

\ConferenceLogo{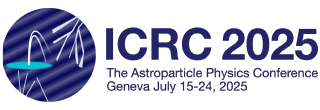}

\FullConference{39th International Cosmic Ray Conference (ICRC2025)\\
 15–24 July 2025\\
Geneva, Switzerland\\}

\begin{document}

\maketitle

\section{Introduction}\label{sec1}

Since their discovery over one hundred years ago, cosmic rays (CR) have been an instrumental tool in broadening our understanding of processes within the Universe - and with good reason: CRs have inprinted on them information on every phenomena they have encountered between their acceleration in their as-of-yet unkown sources and their eventual detection at Earth. A key part of this journey for CRs in the TeV regime is their diffusion through interstellar magnetic fields. In the simplest case we expect the motion of such CRs to be dominated by so-called pitch-angle diffusion in turbulent magnetized plasma, resulting in the arrival direction of CRs being extremely isotropic with the exception of a small dipolar anisotropy \cite{Schlickeiser_2002}. In recent decades, experiments such as IceCube, HAWC, Tibet AS$\gamma$, and others have shown that in addition to these effects, there are also anisotropies across every angular scale able to be probed by these experiments - down to a few degrees \cite{Abbasi_2025, Amenomori_2017,Abeysekara_2019,Bartoli_2018}. The literature on the cause of these smaller-scale anisotropies is extensive, with potential causes ranging from interactions with the heliosphere and more complicated propagation of cosmic rays in the interstellar medium to explanations which require one to go beyond the Standard Model \cite{Ahlers_2017}. Some of these potential causes imply that the Cosmic Ray Anisotropy (CRA) should vary over time \cite{Zhang_2014,Kumar_2019}.

Indeed, the sidereal CRA sky is much more dynamic than one might first think. The Sun and Moon both cast shadows on the Earth leading to a CR deficit. As they travel along the ecliptic, this leads to temporal variations in the sidereal CRA. The former varies with the Solar Cycle, providing a novel way to study the Sun's Magnetosphere over time \cite{Aartsen_2020, Alfaro_2024}. Similarly, the Earth traveling along its orbit sees an induced dipole anisotropy in the direction of its movement due to the Compton-Getting (CG) effect - a dipole whose phase shifts throughout the year in sidereal coordinates, and whose amplitude is proportional to the energy-dependent CR spectral index \cite{Compton_1935}. Although the CG dipole is often used as a means to calibrate CR anisotropy measurements, it has been used to measure the CR spectral index \cite{Amenomori_2008}. Any time-dependence in the CRA offers an exciting way to study physical phenomena - both those that are known, and those which are more speculative

We present a look into the temporal dependence of the TeV CRA utilizing CR muons as measured by the IceCube Neutrino Observatory. This is done in two parts: the first is a model-independent search for time variations in the TeV CRA sky across all angular scales, the second is a measurement of the CG dipole amplitude across several energy bins ranging from a few to hundreds of TeV - and the corresponding measured CR spectral indices. The former offers a methodology by which we can probe a broad number of speculative sources of time variation at once, and the latter demonstrates a novel way by which we can use a well known time variation in the sidereal CRA to probe interesting physics - including allowing us to study the reported bump in the CR spectrum at about 40 TeV \cite{Alemanno_2021,Alfaro_2025,Atkin_2018,Zhang_2021}.

\section{A Model-Indepdent Search for Temporal Dependence in the TeV CRA sky Across All Angular Scales}\label{sec2}

\subsection{Data and Reconstruction}

Here, we utilize a burn-sample of data corresponding to every January from 2012 - 2023, corresponding to $\approx$1/12th of the full dataset available. This is chosen to allow us to search for time variations which occur across a range of time periods. Although we have 31 full calender days per year in this burn sample, we only include full sidereal days, which limits us to 30 days per year. The data selection uses only basic quality cuts, including all events which 1) register in more than 10 of IceCube's Digital Optical Modules (DOMs), 2) Have a reduced Log-Likelihood (RLogL) - a measurement of the event directional reconstrucion goodness of fit - smaller than 25 \cite{Ahrens_2004}, and 3) are recorded when IceCube is taking good data. Although this results in a relatively low median angular resolution of $3.8^\circ$, simulations show us that in this particular model-independent analysis the benefit of larger statistics outweighs potential gains due to a higher angular resolution. This gives us a total of $5.98 \times 10^{10}$ events in our sample. We reconstruct the CRA separately for each sidereal day in our sample using the method described in \cite{Ahlers_2016} with 360 time bins per sidereal day and using a \texttt{HEALPix} sky pixelization of $N_\text{side}=64$ \cite{Gorski_2005} as implemented in the Python library \texttt{Healpy} \cite{Zonca_2019}, corresponding to pixels which are $\approx 0.9^\circ$ across. This reconstruction method has previously been used in the most recent joint IceCube-HAWC measurement of the CRA \cite{Abeysekara_2019}. We take the CRA in equatorial pixel $k$ at time $t$ to be

\begin{equation}
\delta I_{kt} = \frac{n_{kt}}{\langle n_{kt} \rangle} - 1 \, ,
\end{equation}
where $n_{kt}$ and $\langle n_{kt} \rangle$ are the measured counts and the expected isotropic background in pixel $k$ at time $t$, respectively. The latter is obtained from a maximum likelihood estimation. The uncertainty in $\delta I_{kt}$ is calculated from the likelihood function that this reconstruction is maximizing. 

\begin{wrapfigure}{r}{0.5\textwidth}
\includegraphics[width=0.49\textwidth]{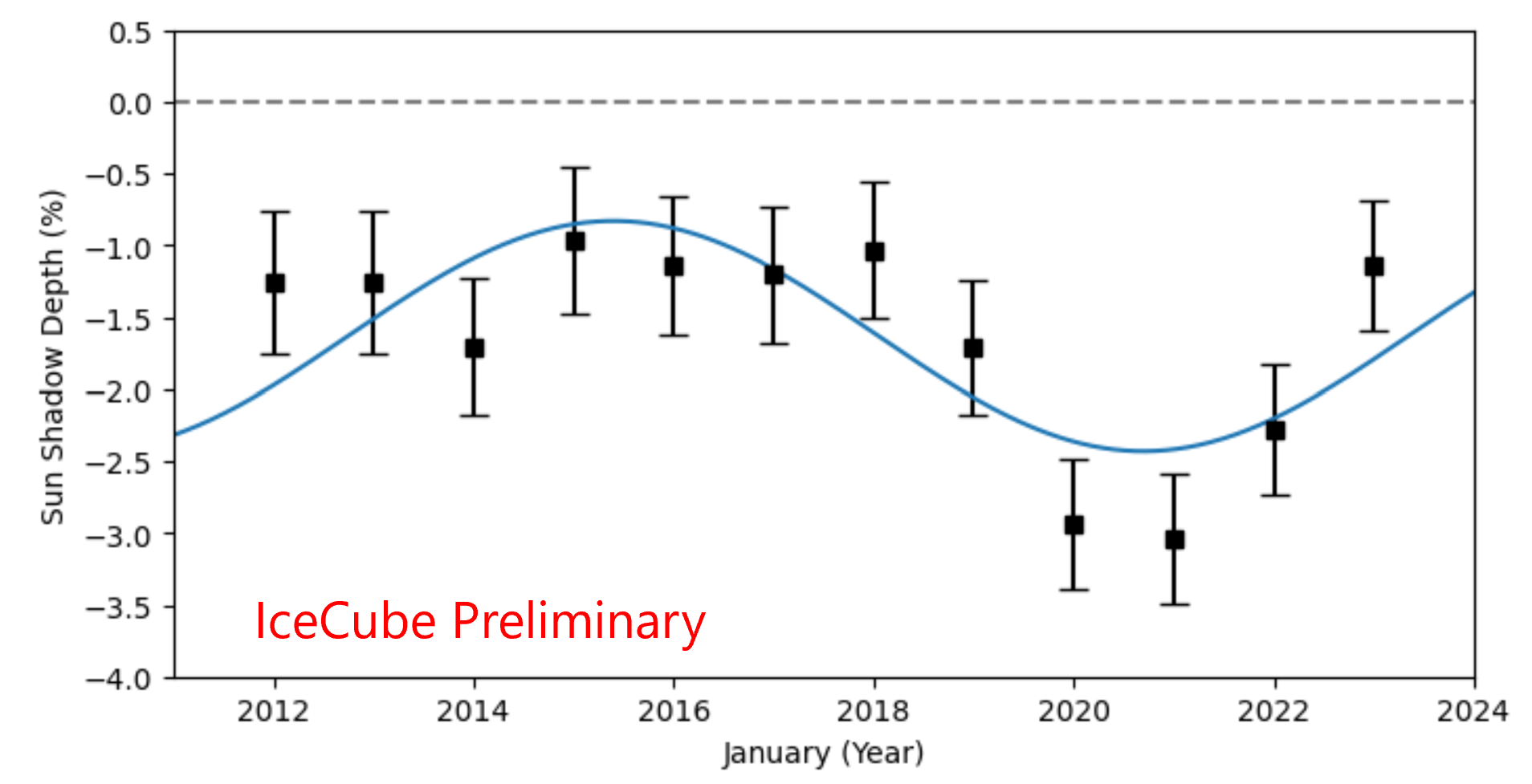}
\caption{Amplitude of the Sun Shadow from 2012 to 2023 in our burn sample (one month per year). A fit cosine is overplotted, with a fit period of $10.6 \pm 1.2$ yr and phase of $140 \pm 40^\circ$ relative to the sun spot minimum associated with the start of solar cycle 24 - consistent with the solar cycle length and expected inverse relationship between sun spot number and Sun Shadow amplitude. The amplitude is attenuated compared to dedicated studies of the Sun Shadow \cite{Aartsen_2020, Alfaro_2024} due to the relatively coarse sky pixelization and low angular resolution.}
\label{fig:sun-shadow}
\end{wrapfigure}

Our day-wise CRA sky-maps contain some known time variations, the most prominent of which is the Sun Shadow. Considering this, we utilize the Sun Shadow as a calibration for our methodology. In previous IceCube CRA analyses, reconstruction was only performed for declinations below $-25^\circ$ latitude due to poor reconstruction near the horizon \cite{Abbasi_2025}. Such a cut would remove the Sun's Shadow from our sky-maps. Due to our use of the Sun Shadow as a calibration source, we perform our reconstruction for declinations below $-10^\circ$ latitude. As a sanity check, we directly search for the Sun Shadow in our CRA sky-maps, and do detect it with a significance of $7.5~\sigma$ with an amplitude that varies with the solar cycle (see Fig. \ref{fig:sun-shadow}). 

\subsection{Methodology}

Taking $n_t$ time bins and $n_k$ pixels, we have a set of $n_t$ \texttt{HEALPix} CRA sky maps, each obtained with data taken during different sidereal days. A model-independent search for time variation takes the form of testing the statistical compatibility of these maps. Doing this, we fit a constant in time for the relative intensity in each pixel, and look at the goodness of this fit via the minimized $\chi^2$. This produces a sky-map of $\chi^2_k$. The null hypothesis we are testing against is that each $\chi^2_k$ is sampled from a $\chi^2$ distribution with $\text{dof}=n_t-1$. We perform complementary tests looking into any deviations from this case. 

\begin{figure}
\centering
    \subfloat{
    \includegraphics[width=0.75\textwidth]{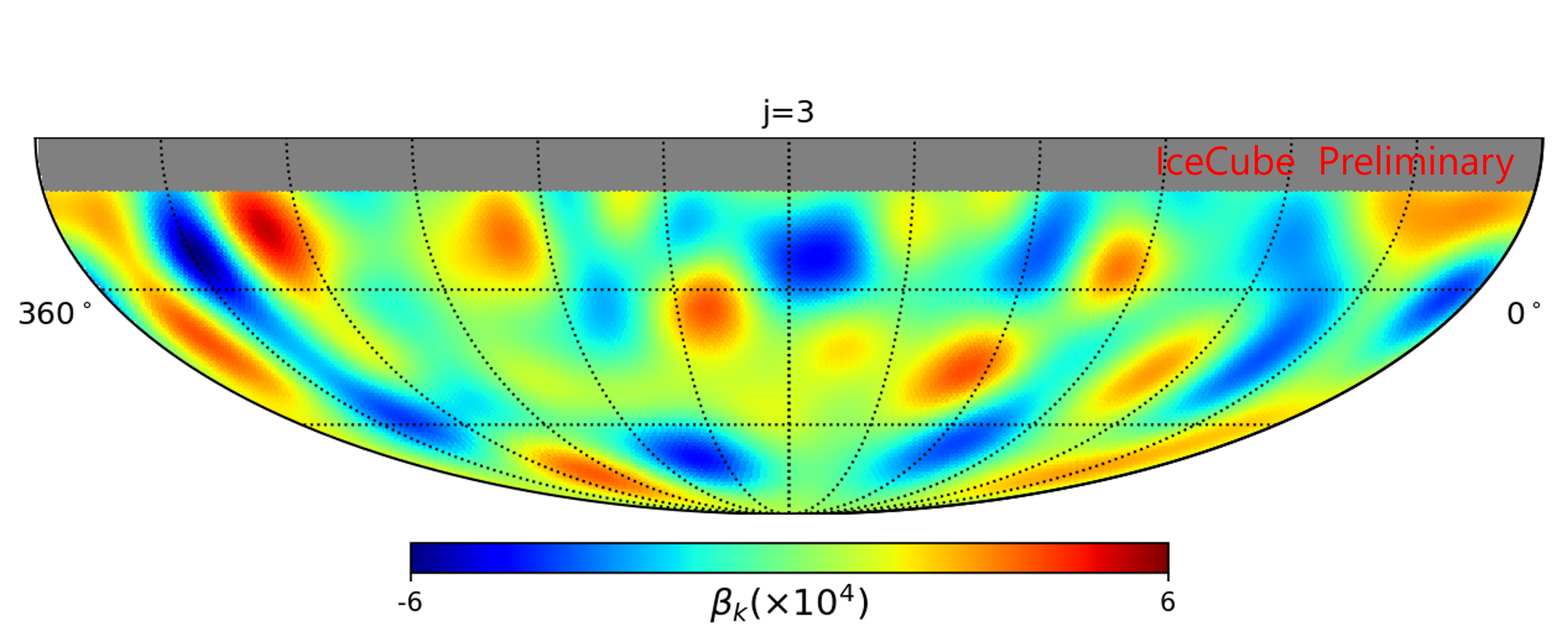}
    }
    
    \subfloat{
    \includegraphics[width=0.32\textwidth]{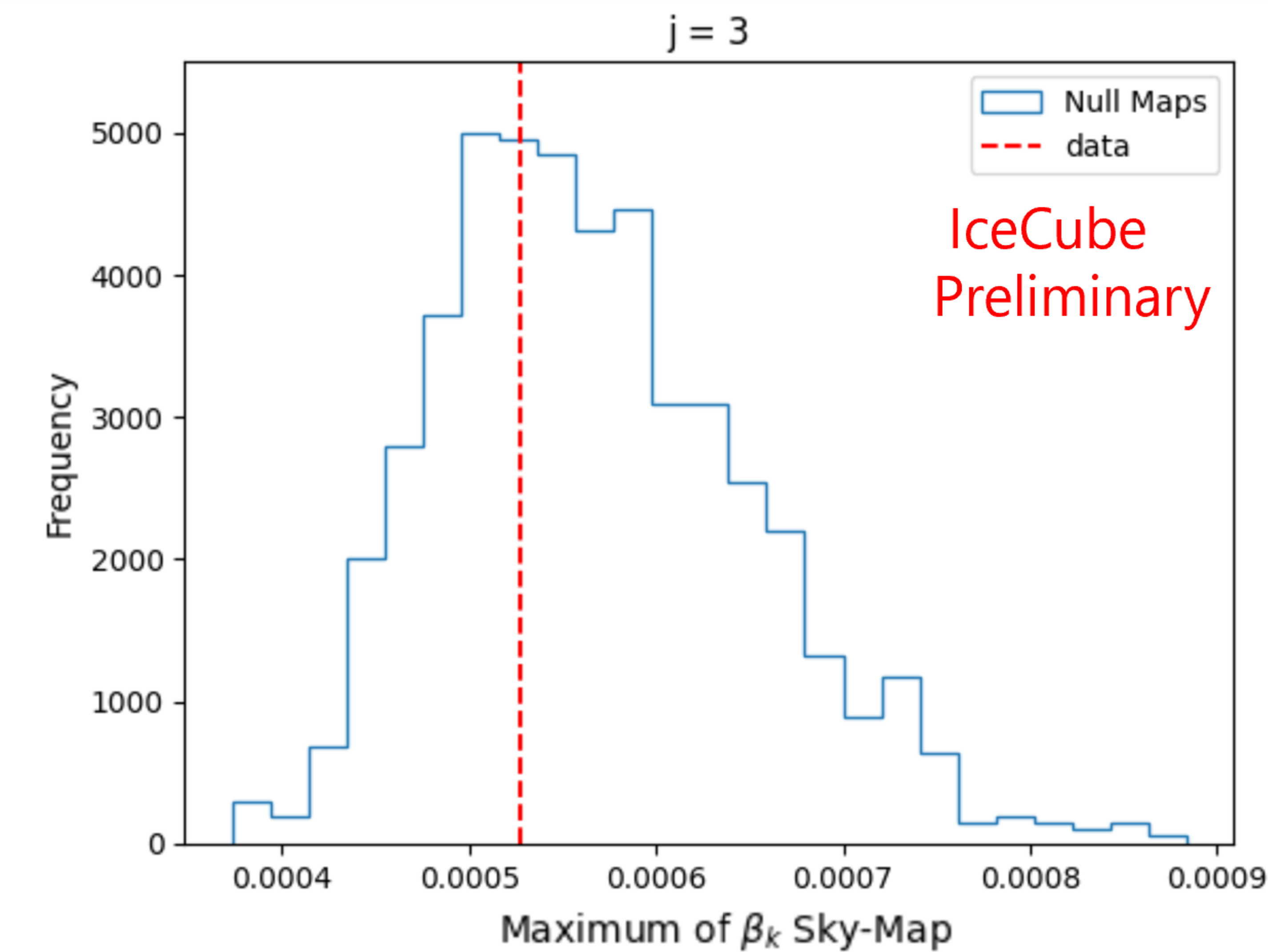}
    \includegraphics[width=0.32\textwidth]{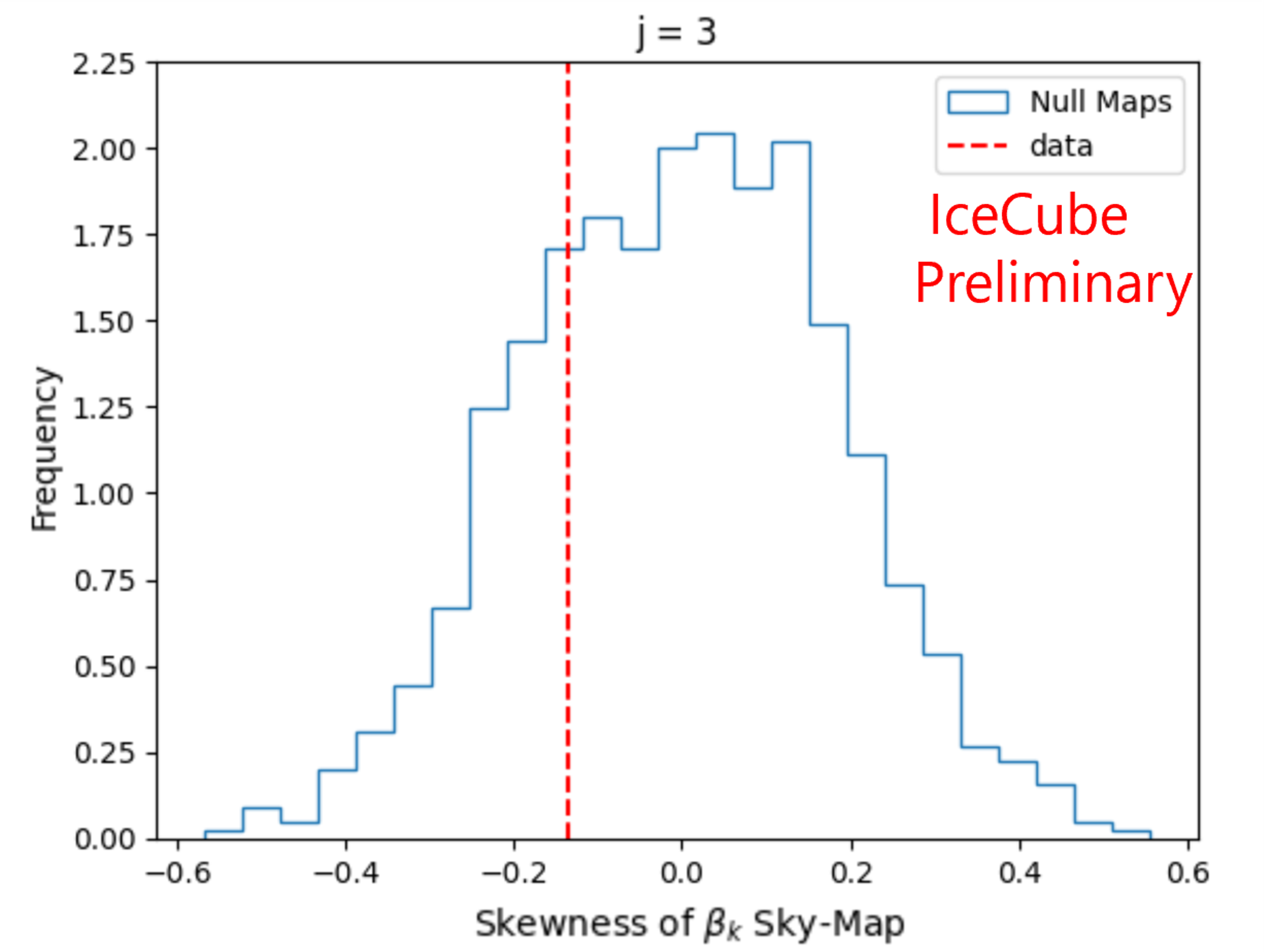}
    \includegraphics[width=0.32\textwidth]{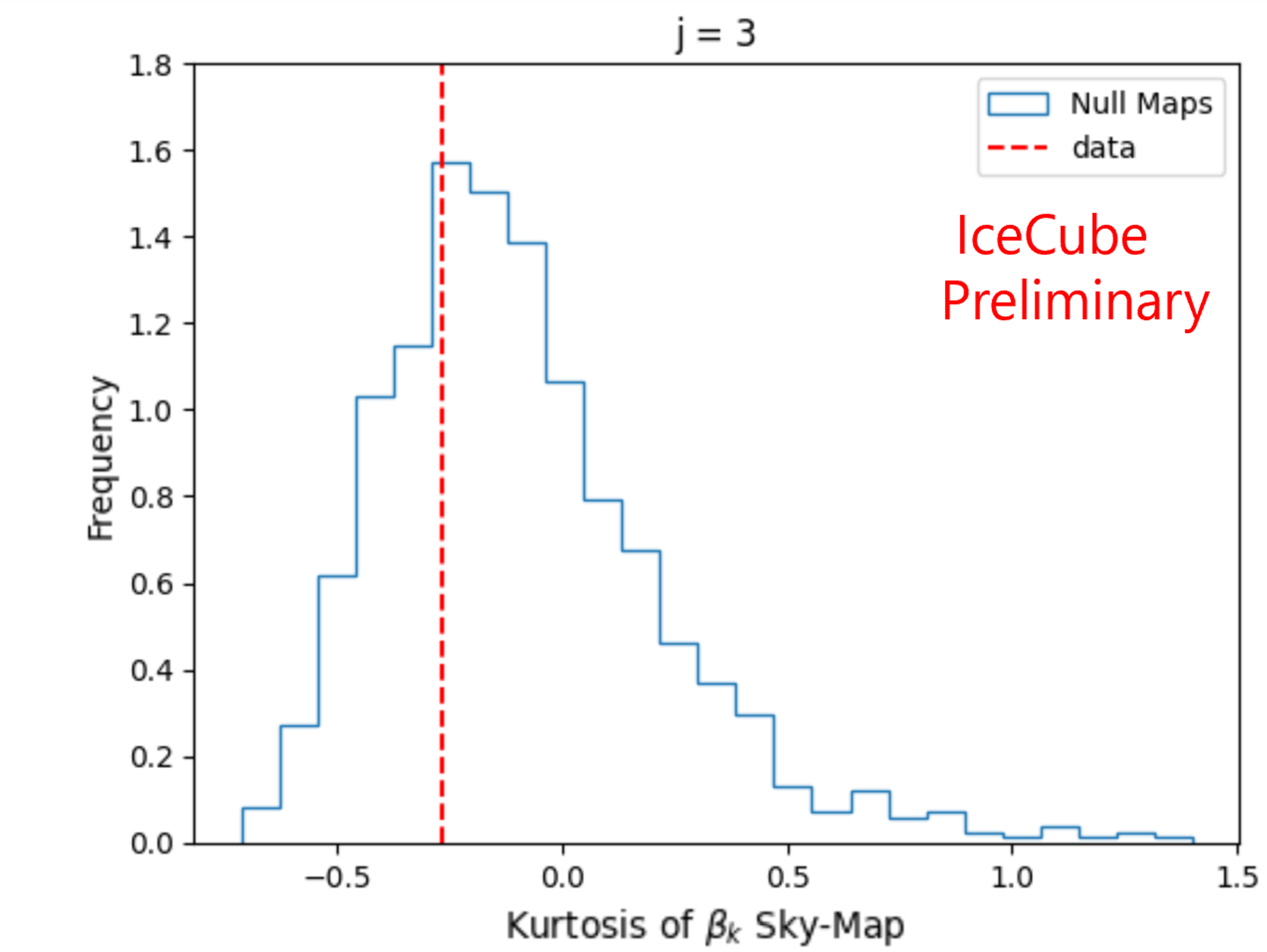}
    }
    \caption{Top: wavelet coefficients $\beta_k$ of the wavelet-smoothed $S_k$ sky-map with a Mexican needlet with angular scale parameter $j=3$. Bottom: maximum $\beta_k$ (left), skewness of the distribution of $\beta_k$ (middle), kurtosis of the distribution of $\beta_k$ (right) of the above sky-map compared against the respective distributions each is sampled from, assuming the null case.}
    \label{fig:needlet_coeff}
\end{figure}

\begin{figure}
    \includegraphics[width=0.49\textwidth]{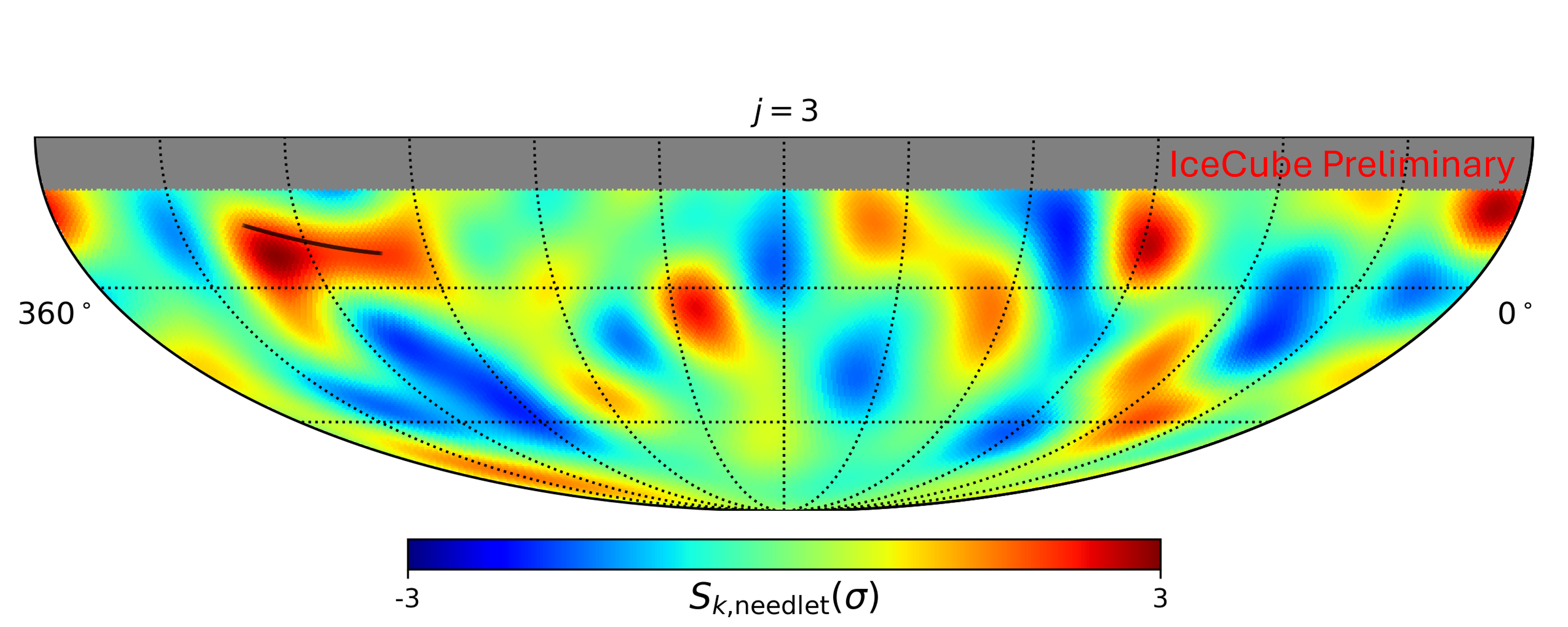}
    \includegraphics[width=0.49\textwidth]{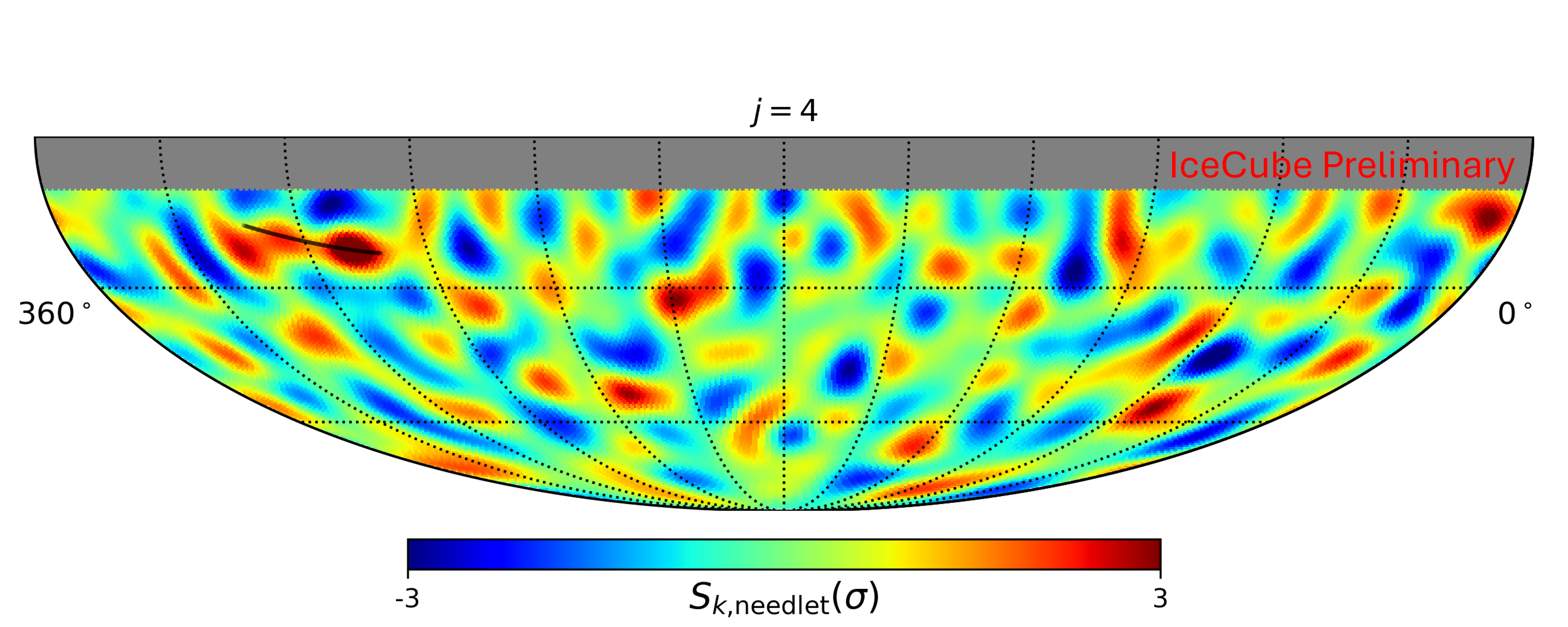}
    \caption{Local significance of the $j=3$ (left) and $j=4$ (right) Mexican needlet coefficients obtained from phase-folded CRA sky-maps with a period of one year, calculated using the method described in \cite{Aab_2017}. The path of the sun in our data set is overplotted in black.}
    \label{fig:phase-fold}
\end{figure}

The simplest test we implement is calculating $\chi^2_\text{full-sky} = \sum\limits_i\chi^2_k$. Under the null hypothesis the resulting $\chi^2_\text{full-sky}$ will be sampled from a $\chi^2$ distribution with $\text{dof}=n_k(n_t-1)$, allowing us to use it as a simple full-sky test statistic. 

Additionally, we perform tests to look for deviations which are localized on the sky. These tests rely on decomposing our map in terms of spherical harmonics. Due to IceCube's incomplete sky coverage, there is mixing between multipole modes~\citep{SOMMERS2001271, Abeysekara_2019}. This is a particular problem for a sky-map of $\chi^2$ values due to the strong monopole term, which after mixing will drown out any potential signal at larger multipoles. To alleviate this, we transform our $\chi^2_k$ sky-map into a local significance sky-map $S_k$. We begin by calculating the p-value ($p_k$) associated with each $\chi^2_k$, and then calculating the pixel-wise local significance from each p-value via 
\begin{equation}
S_k = \sqrt{2} \text{ erf}^{-1}(1-2p_k) \, .
\end{equation}
Under the null hypothesis, the $S_k$ would follow a standard normal distribution \cite{bohm_2017} resulting in a sky-map with a flat pseudo-$C_\ell$ ($\hat{C_\ell}$) power spectrum. Using this new sky-map, we perform two local and scale-dependent analyses, largely inspired by \cite{Aab_2017}. The first of these is simply analyzing the power spectrum of our $S_k$ sky-map and looking for deviations from the null case, which we do via \texttt{Healpy}'s \texttt{anafast} routine. The $\hat{C_\ell}$ expected in the null case is estimated by creating sky-maps populated with values sampled from a standard normal distribution, masking regions above $-10^\circ$ dec (as is done with the data) and performing the above analysis on each of the maps. 

In the second analysis, we convolve this map with Mexican Needlets --- a type of wavelet --- which are chosen due to their strong localization in real space \cite{Scodeller_2011}. These wavelets have two user-defined parameters, $B \in \mathbb{R}^+$ and $p \in \mathbb{Z}^+$, controlling their shape --- once these are set, we are provided with a family of kernels probing a discrete set of angular scales, numbered by $j \in \mathbb{Z}$. We set $p=1$ because it has maximum localization in real space. Due to these needlets probing discrete sets of angular scales, $B$ can be determined by deciding on a) the minimum and maximum angular scales we would like to probe and b) the number of angular scales between them that we would like to consider. Probing from $180^\circ-3^\circ$ with five angular scales is well satisfied by taking $B=2$ and looking at $j=1-5$ (opting not to look at $j=0$ due to it being dominated by the dipole, which is already considered in the power spectrum analysis). This convolution produces a smoothed sky-map, in which each pixel $k$ contains a so-called wavelet coefficient $\beta_k$.

\begin{wrapfigure}{r}{0.5\textwidth}
\includegraphics[width=0.49\textwidth]{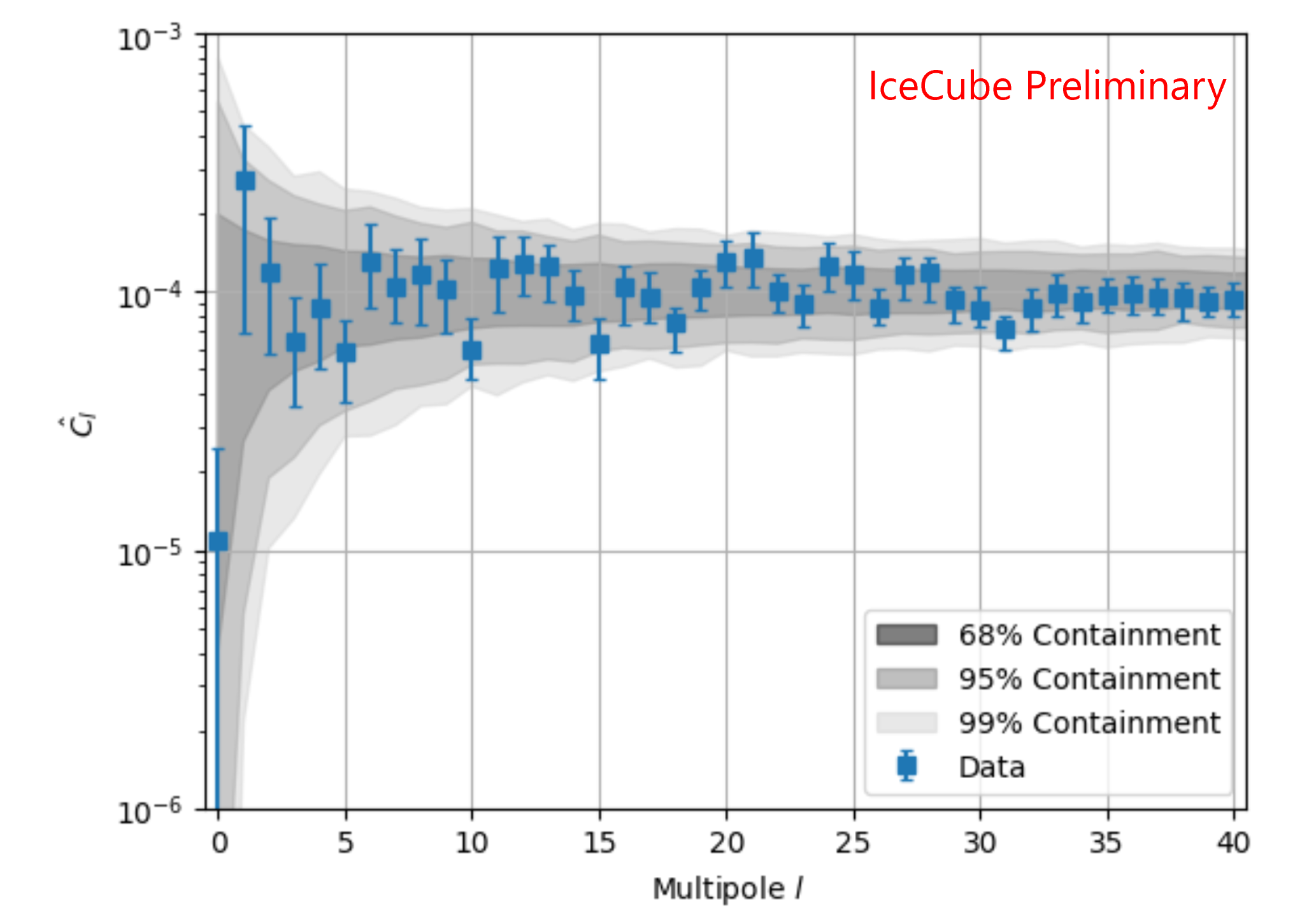}
\caption{$\hat{C_\ell}$ power spectrum of the local significance map $S_k$ described in the text. Uncertainties on the $\hat{C_\ell}$ are systematic and calculated in the same as done in \cite{Abbasi_2025}.}
\label{fig:powerspec}
\end{wrapfigure}

In order to calculate deviations of our wavelet-smoothed sky-maps from the null case, we consider three test statistics for each $\beta_k$ sky-map: the maximum $\beta_k$, the skewness of the distribution of $\beta_k$, and the kurtosis of the distribution of $\beta_k$. The distribution of each statistic expected in the null case is found in a way analogous to that done in the power spectrum analysis (see Fig. \ref{fig:needlet_coeff} for an example of this analysis with our $j=3$ wavelet). Note that in typical wavelet analyses (\cite{Ade_2016}, for example), regions near masked regions are not considered after a wavelet convolution due to the needlet coefficients being contaminated. However, since we are interested in regions near the horizon, we do not do this. Any potential contamination by our masked regions is already taken into account when obtaining the statistical distribution expected in the null case.

\begin{figure}[h!]
\centering
   
    \includegraphics[width=0.32\textwidth]{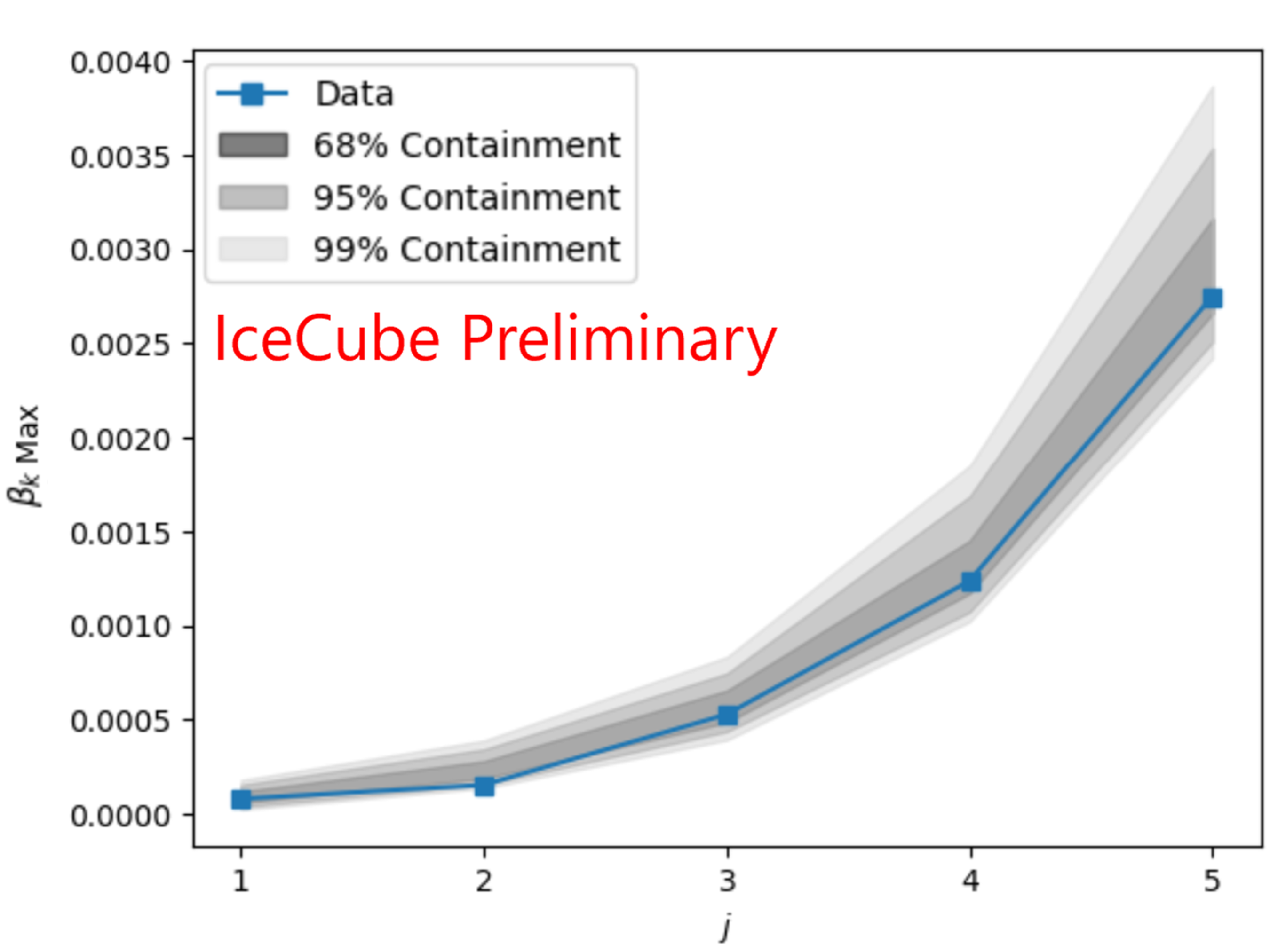}
    \includegraphics[width=0.32\textwidth]{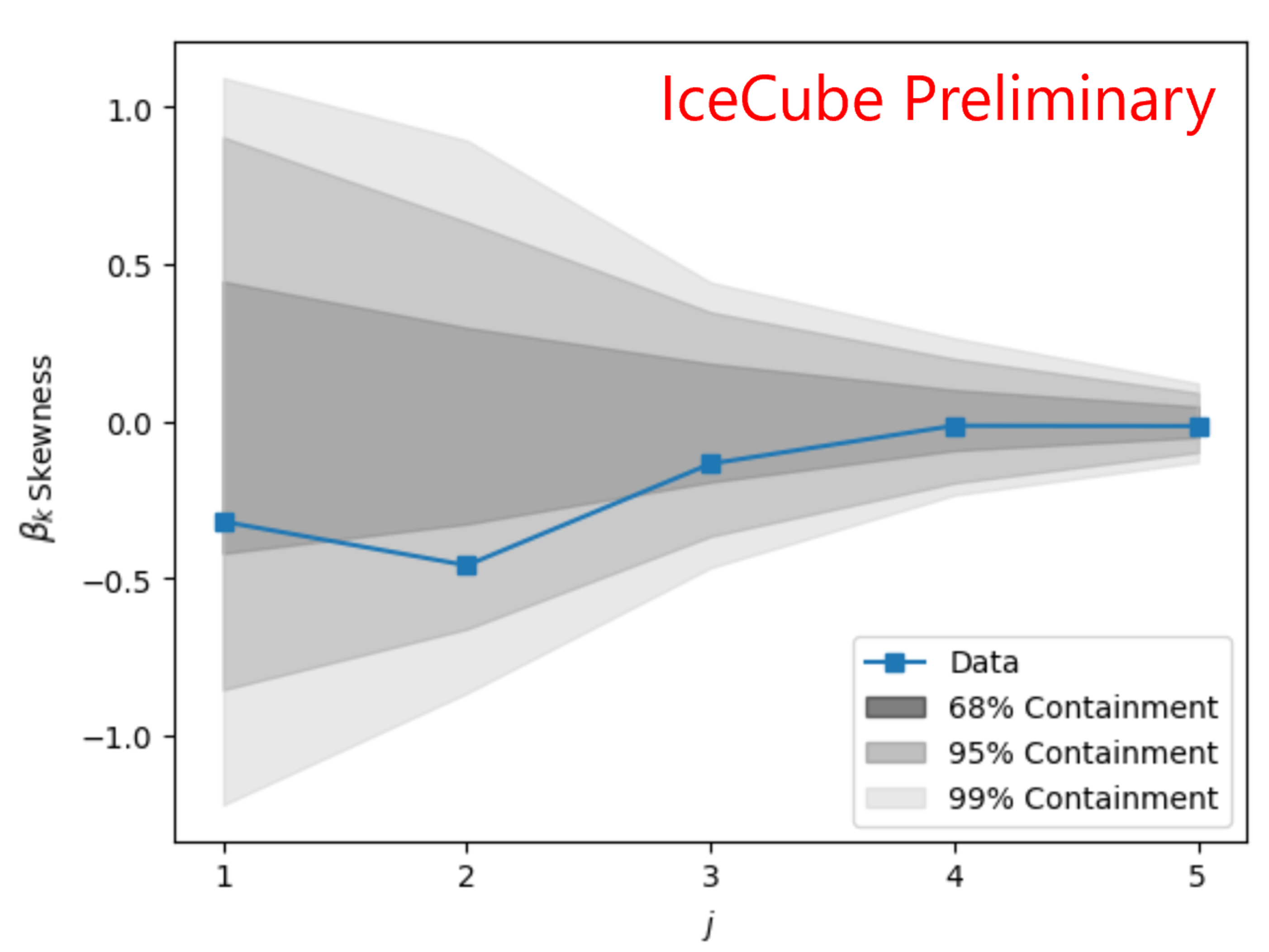}
    \includegraphics[width=0.32\textwidth]{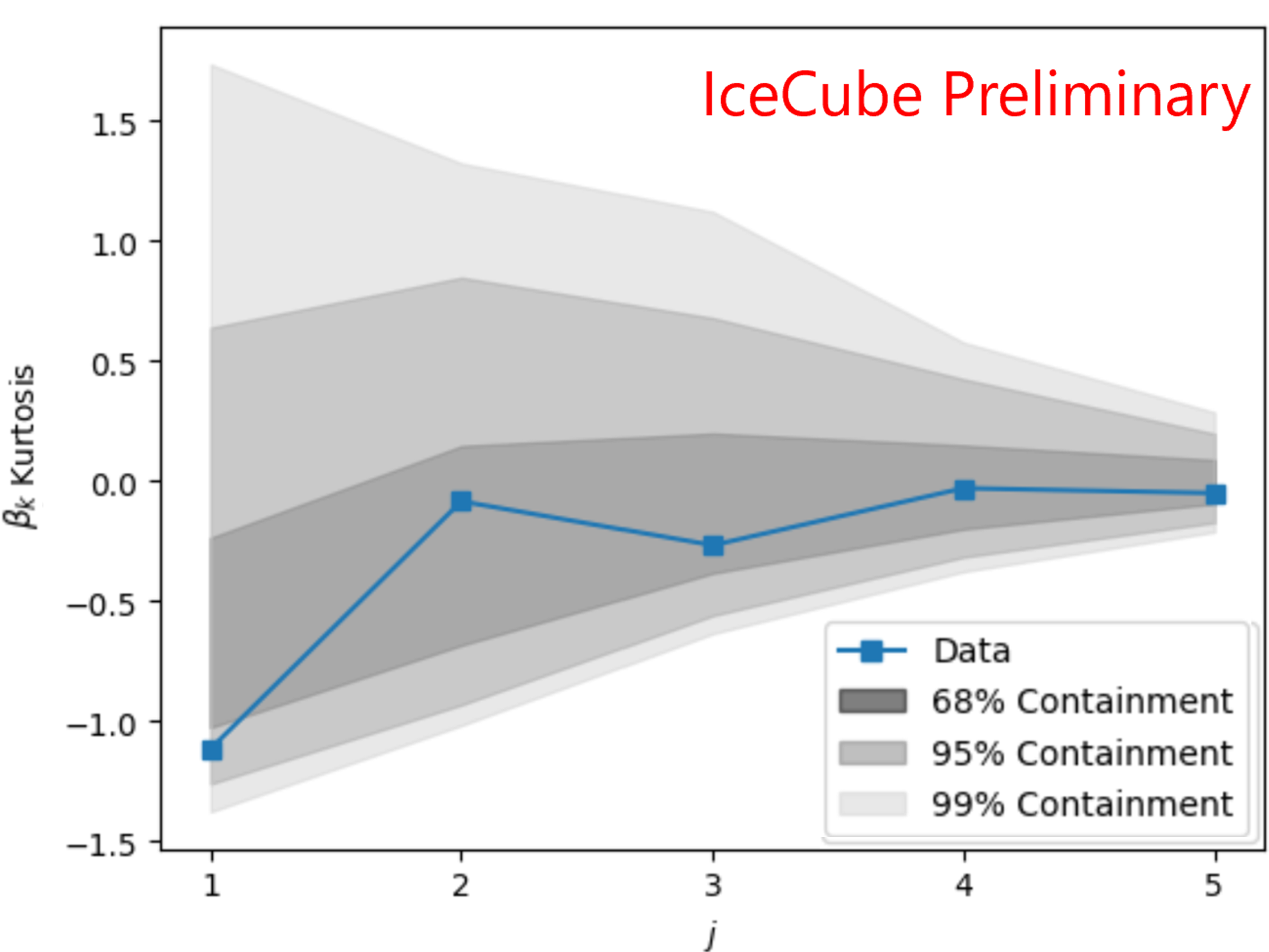}
    \caption{Needlet test statistics for the five needlets used in this study compared against the spread expected in the null case, including the maximum $\beta_k$ (left), skewness of the distribution of $\beta_k$ (middle), and kurtosis of the distribution of $\beta_k$ (right).}
    \label{fig:needlet_test}
\end{figure}

\subsection{Results and Discussion}

Performing the full-sky analysis, we find $\frac{\chi^2}{\text{ndof}} = \frac{155656}{156288}$ corresponding to a p-value of 0.36, and so no significant deviation from the null hypothesis is found in this test. No significant deviations are found when looking at the needlet test statistics (see Fig. \ref{fig:needlet_test}). However, we can see a somewhat prominent (albeit not statistically significant) region just below $-10^\circ$ declination when looking at the $j = 3$ needlet-smoothed map (Fig. \ref{fig:needlet_coeff}), which is in line with where we expect the Sun Shadow to be in our data set. Probing this further, we phase-fold our CRA sky-maps with a period of one year to emphasize any time variations with this period. This is done by averaging sky-maps of sidereal days which are one year apart, giving us 30 CRA sky-maps. Performing the same analysis and examining the resulting local significance sky-maps, we observe that this region is larger and more significant, indicating it is likely the Sun Shadow (Fig. \ref{fig:phase-fold}). We have performed subsequent sensitivity studies by injecting an artificial Sun Shadow into artificial data and performing the analysis described above. We find the results we see are consistent with how we would expect the Sun Shadow to appear in this analysis. No significant deviations are found in the power spectrum analysis (see Fig. \ref{fig:powerspec}).

\section{Measurement of the All-Particle CR Spectral Index from 10 TeV to 300 TeV using the CG Effect}\label{sec3}
The first--order Compton--Getting effect~\cite{Compton_1935} predicts a dipole anisotropy in the form
\begin{equation}
\delta I= 
\frac{v(t)}{c}\left[\gamma(E)+2\right]\cos{\xi}\,,
\end{equation}\label{eq:compton_getting}
where $I$ is the cosmic ray intensity, $\gamma(E)$ the cosmic--ray spectral index, $v(t)/c$ the ratio of the observer's velocity --in this case Earth’s orbital velocity-- to the speed of light, and $\xi$ the angle between the cosmic ray particle’s arrival direction and the direction of Earth's motion. Of note is that the spectral index $\gamma(E)$ depends on the energy $E$. This suggests an alternative method for statistically inferring the shape of the energy spectrum using the arrival direction distribution of cosmic rays.

\subsection{Data and Reconstruction}

Unlike the previous analysis, this measurement shows an increased sensitivity with better angular resolution. Striking a balance between this and the larger statistics one gets when including worse-reconstructed events, we follow previous Moon and Sun Shadow analyses with IceCube \cite{Aartsen_2014} and maximize $\frac{\sqrt{\eta}}{\Delta \psi_\text{med}}$ as a function of RLogL, where $\eta$ is the number of events passing the cut and  $\Delta \psi_\text{med}$ is the resulting median angular resolution. This is optimized at RLogL = 13.3, giving a median angular resolution of $3^\circ$. We cut on this, as well as only including events which trigger 10 or more DOMs, and events which are recorded when IceCube is taking good data. We use data taken during the period between May 2011 and May 2012, corresponding to $\approx$1/12th of the whole dataset. 
Unlike the previous analysis, this analysis benefits from a full year of data because the signal from the sidereal reference cancels out after rotating $360^\circ$~\cite{DiazVelez:2021zT}. As with the previous analysis, events are binned using a \texttt{HEALPix} sky pixelization of $N_\text{side}=64$ \cite{Gorski_2005} as implemented in the Python library \texttt{Healpy}\cite{Zonca_2019} in a solar reference frame. Events are binned into nine energy bins are separated by $\Delta\log_{10}{E_\mathrm{Reco}}= 0.25$. The energy is determined statistically from the number of triggered optical modules and the cosine of the reconstructed zenith angle $\theta$. This results in median energies that are roughly equidistant logarithmically. For more details on the energy estimation see Ref.~\cite{Abbasi_2025}. Due to low statistics at higher energies, we only utilize the first eight of these energy bins. Coupled with the data cuts, this gives us $3.52 \times 10^{10}$ events in this sample.

\subsection{Analysis and Results}

The method for estimating the background requires averaging along each declination band, thereby eliminating the vertical dependency (i.e., the $m=0$ spherical harmonics) ~\cite{Abbasi_2025}.
\begin{wrapfigure}{r}{0.5\textwidth}
\includegraphics[width=0.49\textwidth]{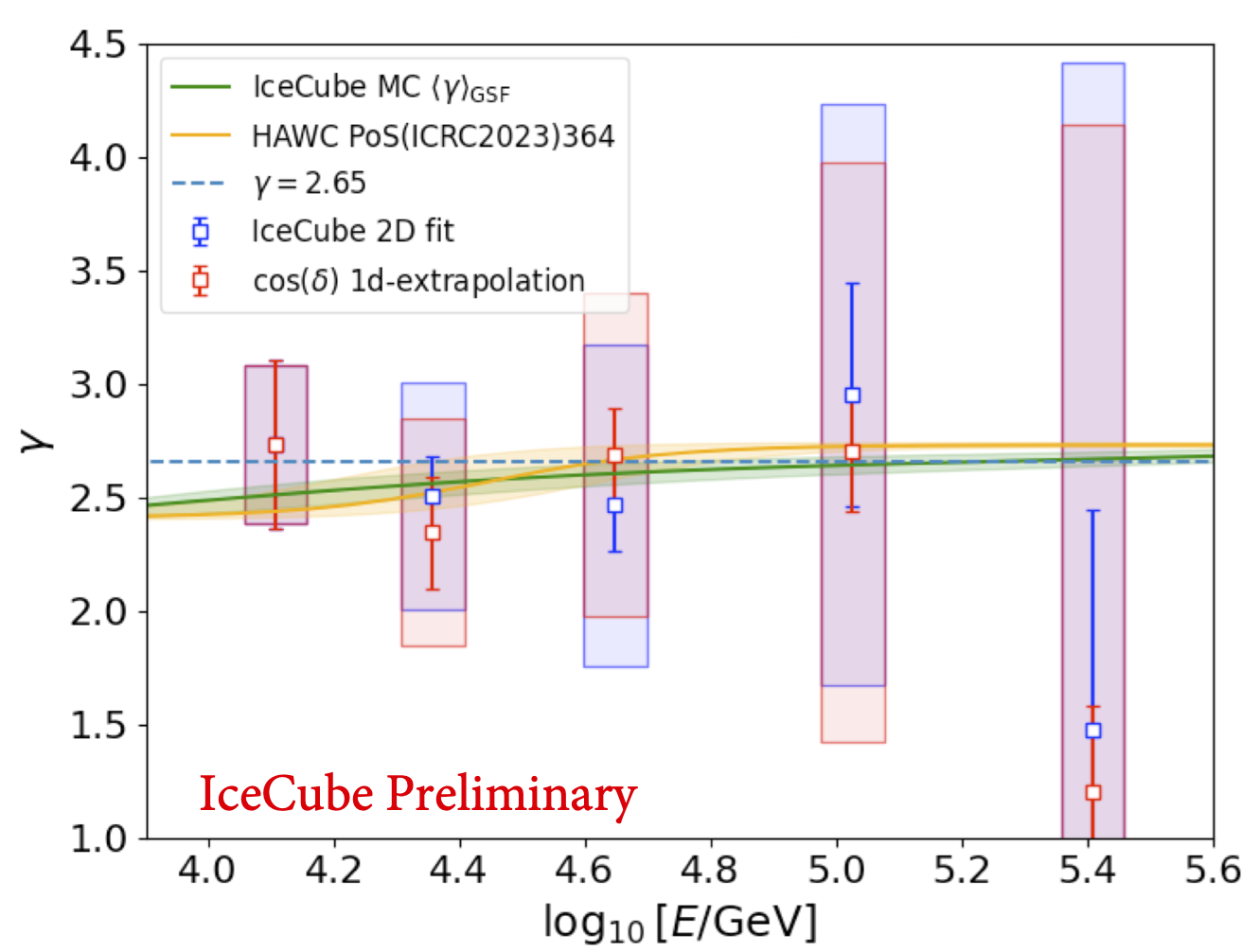}
\caption{Our measurement of the all-particle spectral index $\gamma$ compared against the three representative spectra using the two measurement methods described in the text. The error bars are statistical, and the boxes are systematics - principally due to interference from the extended-sidereal frame and the variation of Earth's orbital speed through the year.}
\label{fig:solar-dipole_meas}
\end{wrapfigure}
Taking this into account, a dipole with a phase of $\phi$ and amplitude $A$ measured at RA $\alpha$ and dec $\delta$ takes the form
\begin{equation}
\delta I_\text{dip}(\alpha, \delta) = A\cos(\delta)\cos(\alpha-\phi)\,. \label{eq:CG}
\end{equation}
Alternatively, we can use the 1-dimensional projection of the dipole $A_{\delta_i}\cos(\alpha-\phi)$.
Since the amplitude of the projected dipole decreases with declination away from the equator by $\cos{\delta}$, we can measure the 1-dimensional dipole amplitude for different declination bands and fit a cosine function to these measurements, extrapolating to the equator. Finally, we obtain a value of $\gamma$ from Eq.~\ref{eq:compton_getting} and the dipole amplitude $\delta I$ from both methods. 

We would like to test whether our measurements are consistent with three spectra: 1) one with a constant $\gamma(E) = -2.65$, 2) A spectrum recently published by HAWC, containing a break at about 30 TeV \cite{Morales-Soto_2023}, and 3) The expected response from IceCube to the HAWC spectrum.
To appropriately compare the measured spectral indices with expected values, we compare to the mean spectral index in each of our energy bins, weighted by their energy distributions $N(E)$:
\begin{equation}
\langle \gamma \rangle = \frac{\int N(E) \gamma(E) dE}{\int N(E)dE}\,.
\end{equation}
The comparison against the predicted spectra is done using a $\chi^2$ test.
The dipole measurements of the burn sample above 250 TeV are not significant so we compare the measurements of only the first 5 bins which are determined to be compatible with all representative spectra, though it is clear that this is driven by the large error bars (see Fig \ref{fig:solar-dipole_meas}). 

\section{Conclusion and Outlook}\label{sec4}

We have performed two studies of the CRA using burn samples of IceCube CR muons - a model-independent search for time variation in the CRA, and a measurement of the all-particle CR spectral index in the TeV regime. In the first, we described a novel method to look for time variation in the TeV CRA sky, assuming no initial hypothesis. We detected no significant signals, but do see a relatively prominent region coincident with the expected Sun Shadow. In the second, we performed a measurement of all-particle CR spectral index utilizing the CG effect. These measurements were consistent with expectations, but were unable to distinguish between different spectral models. We expect both studies to improve when using the full IceCube dataset, which offers an order of magnitude more events.

\scriptsize{
\providecommand{\href}[2]{#2}\begingroup\raggedright\endgroup


\begin{thebibliography}{10}

\bibitem{Schlickeiser_2002}
R.~Schlickeiser, \href{http://dx.doi.org/10.1007/978-3-662-04814-6}{{\em Cosmic Ray Astrophysics}}.
\newblock Springer, 2002.

\bibitem{Abbasi_2025}
{\bfseries IceCube} Collaboration, R.~Abbasi {\em et~al.}, \href{http://dx.doi.org/10.3847/1538-4357/adb1de}{{\em ApJ} {\bfseries 981} no.~2, (Mar, 2025) 182}.

\bibitem{Amenomori_2017}
{\bfseries Tibet AS$\gamma$} Collaboration, M.~Amenomori {\em et~al.}, \href{http://dx.doi.org/10.3847/1538-4357/836/2/153}{{\em ApJ} {\bfseries 836} no.~2, (2017) 153}.

\bibitem{Abeysekara_2019}
{\bfseries IceCube and HAWC} Collaboration, A.~U. Abeysekara {\em et~al.}, \href{http://dx.doi.org/10.3847/1538-4357/aaf5cc}{{\em ApJ} {\bfseries 871} no.~1, (2019) 96}.

\bibitem{Bartoli_2018}
{\bfseries ARGO-YBJ} Collaboration, B.~Bartoli {\em et~al.}, \href{http://dx.doi.org/10.3847/1538-4357/aac6cc}{{\em Astrophys. J.} {\bfseries 861} no.~2, (2018) 93}.

\bibitem{Ahlers_2017}
M.~{Ahlers} and P.~{Mertsch}, \href{http://dx.doi.org/10.1016/j.ppnp.2017.01.004}{{\em Prog. Part. Nucl. Phys.} {\bfseries 94} (May, 2017) 184--216}.

\bibitem{Zhang_2014}
M.~Zhang, P.~Zuo, and N.~Pogorelov, \href{http://dx.doi.org/10.1088/0004-637X/790/1/5}{{\em ApJ} {\bfseries 790} no.~1, (Jun, 2014) 5}.

\bibitem{Kumar_2019}
R.~Kumar {\em et~al.}, \href{http://dx.doi.org/10.1093/mnras/sty3141}{{\em Mon. Not. R. Astron. Soc.} {\bfseries 483} no.~1, (Feb., 2019) 896–900}.

\bibitem{Aartsen_2020}
{\bfseries IceCube} Collaboration, M.~G. Aartsen {\em et~al.}, \href{http://dx.doi.org/10.1103/physrevd.103.042005}{{\em Phys. Rev. D} {\bfseries 103} no.~4, (2020) 042005}.

\bibitem{Alfaro_2024}
{\bfseries HAWC} Collaboration, R.~Alfaro {\em et~al.}, \href{http://dx.doi.org/10.3847/1538-4357/ad3208}{{\em ApJ} {\bfseries 966} no.~1, (2024) 67}.

\bibitem{Compton_1935}
A.~H. Compton and I.~A. Getting, \href{http://dx.doi.org/10.1103/PhysRev.47.817}{{\em Phys. Rev.} {\bfseries 47} (Jun, 1935) 817--821}.

\bibitem{Amenomori_2008}
{\bfseries Tibet AS$\gamma$} Collaboration, M.~Amenomori {\em et~al.}, \href{http://dx.doi.org/10.1016/j.nuclphysbps.2007.11.044}{{\em Nucl. Phys. B} {\bfseries 175} (2008) 427–430}.

\bibitem{Alemanno_2021}
{\bfseries DAMPE} Collaboration, F.~{Alemanno} {\em et~al.}, \href{http://dx.doi.org/10.1103/PhysRevLett.126.201102}{{\em Phys. Rev. Lett.} {\bfseries 126} no.~20, (May, 2021) 201102}.

\bibitem{Alfaro_2025}
{\bfseries HAWC} Collaboration, R.~Alfaro {\em et~al.}, \href{http://dx.doi.org/10.1016/j.astropartphys.2024.103077}{{\em Astropart. Phys.} {\bfseries 167} (May, 2025) 103077}.

\bibitem{Atkin_2018}
{\bfseries NUCLEON} Collaboration, E.~Atkin {\em et~al.}, \href{http://dx.doi.org/10.1134/s0021364018130015}{{\em JETP Letters} {\bfseries 108} no.~1, (2018) 5–12}.

\bibitem{Zhang_2021}
P.~Zhang {\em et~al.}, \href{http://dx.doi.org/10.1088/1475-7516/2021/05/012}{{\em JCAP} {\bfseries 2021} no.~05, (2021) 012}.

\bibitem{Ahrens_2004}
{\bfseries AMANDA} Collaboration, J.~Ahrens {\em et~al.}, \href{http://dx.doi.org/https://doi.org/10.1016/j.nima.2004.01.065}{{\em Nucl.\ Inst.\ Meth.\ A} {\bfseries 524} no.~1, (2004) 169--194}.

\bibitem{Ahlers_2016}
M.~Ahlers {\em et~al.}, \href{http://dx.doi.org/10.3847/0004-637X/823/1/10}{{\em ApJ} {\bfseries 823} no.~1, (May, 2016) 10}.

\bibitem{Gorski_2005}
K.~M. {G{\'o}rski} {\em et~al.}, \href{http://dx.doi.org/10.1086/427976}{{\em ApJ} {\bfseries 622} no.~2, (Apr., 2005) 759--771}.

\bibitem{Zonca_2019}
A.~Zonca {\em et~al.}, \href{http://dx.doi.org/10.21105/joss.01298}{{\em JOSS} {\bfseries 4} no.~35, (2019) 1298}.

\bibitem{Aab_2017}
{\bfseries Pierre-Auger} Collaboration, A.~Aab {\em et~al.}, \href{http://dx.doi.org/10.1088/1475-7516/2017/06/026}{{\em JCAP} {\bfseries 2017} no.~06, (Jun, 2017) 026}.

\bibitem{SOMMERS2001271}
P.~Sommers, \href{http://dx.doi.org/https://doi.org/10.1016/S0927-6505(00)00130-4}{{\em Astropart. Phys.} {\bfseries 14} no.~4, (2001) 271 -- 286}.

\bibitem{bohm_2017}
G.~Bohm and G.~Zech, \href{http://dx.doi.org/10.3204/pubdb-2017-08987}{{\em Introduction to Statistics and Data Analysis for Physicists, Third Revised Edition}}.
\newblock Verlag Deutsches Elektronen-Synchrotron, 2017.

\bibitem{Scodeller_2011}
S.~Scodeller {\em et~al.}, \href{http://dx.doi.org/10.1088/0004-637X/733/2/121}{{\em ApJ} {\bfseries 733} no.~2, (May, 2011) 121}.

\bibitem{Ade_2016}
{\bfseries Planck} Collaboration, P.~A.~R. Ade {\em et~al.}, \href{http://dx.doi.org/10.1051/0004-6361/201526681}{{\em A\&A} {\bfseries 594} (Sept., 2016) A16}.

\bibitem{Aartsen_2014}
{\bfseries IceCube} Collaboration, M.~G. Aartsen {\em et~al.}, \href{http://dx.doi.org/10.1103/PhysRevD.89.102004}{{\em Phys. Rev. D} {\bfseries 89} (May, 2014) 102004}.

\bibitem{DiazVelez:2021zT}
J.~C. D\'iaz~V\'elez, P.~Desiati, R.~Abbasi, and F.~McNally, \href{http://dx.doi.org/10.22323/1.395.0085}{{\em PoS} {\bfseries ICRC2021} (2021) 085}.

\bibitem{Morales-Soto_2023}
J.~A. Morales-Soto and J.~C. Arteaga-Velázquez, \href{http://dx.doi.org/10.22323/1.444.0364}{{\em PoS} {\bfseries ICRC2023} (2023) 364}.

\end{thebibliography}
}

\clearpage

\section*{Full Author List: IceCube Collaboration}

\scriptsize
\noindent
R. Abbasi$^{16}$,
M. Ackermann$^{63}$,
J. Adams$^{17}$,
S. K. Agarwalla$^{39,\: {\rm a}}$,
J. A. Aguilar$^{10}$,
M. Ahlers$^{21}$,
J.M. Alameddine$^{22}$,
S. Ali$^{35}$,
N. M. Amin$^{43}$,
K. Andeen$^{41}$
% ...
\\
\\
% ...
$^{10}$ Universit{\'e} Libre de Bruxelles, Science Faculty CP230, B-1050 Brussels, Belgium \\
% ...
$^{16}$ Department of Physics, Loyola University Chicago, Chicago, IL 60660, USA \\
$^{17}$ Dept. of Physics and Astronomy, University of Canterbury, Private Bag 4800, Christchurch, New Zealand \\
% ...
$^{21}$ Niels Bohr Institute, University of Copenhagen, DK-2100 Copenhagen, Denmark \\
$^{22}$ Dept. of Physics, TU Dortmund University, D-44221 Dortmund, Germany \\
% ...
$^{35}$ Dept. of Physics and Astronomy, University of Kansas, Lawrence, KS 66045, USA \\
% ...
$^{39}$ Dept. of Physics and Wisconsin IceCube Particle Astrophysics Center, University of Wisconsin{\textemdash}Madison, Madison, WI 53706, USA \\
% ...
$^{41}$ Department of Physics, Marquette University, Milwaukee, WI 53201, USA \\
% ...
$^{43}$ Bartol Research Institute and Dept. of Physics and Astronomy, University of Delaware, Newark, DE 19716, USA \\
% ...
$^{63}$ Deutsches Elektronen-Synchrotron DESY, Platanenallee 6, D-15738 Zeuthen, Germany \\
$^{\rm a}$ also at Institute of Physics, Sachivalaya Marg, Sainik School Post, Bhubaneswar 751005, India \\
% ...

\subsection*{Acknowledgments}

\noindent
The authors gratefully acknowledge the support from the following agencies and institutions:
USA {\textendash} U.S. National Science Foundation-Office of Polar Programs,
U.S. National Science Foundation-Physics Division,
U.S. National Science Foundation-EPSCoR,
U.S. National Science Foundation-Office of Advanced Cyberinfrastructure,
Wisconsin Alumni Research Foundation,
Center for High Throughput Computing (CHTC) at the University of Wisconsin{\textendash}Madison,
Open Science Grid (OSG),
Partnership to Advance Throughput Computing (PATh),
Advanced Cyberinfrastructure Coordination Ecosystem: Services {\&} Support (ACCESS),
Frontera and Ranch computing project at the Texas Advanced Computing Center,
U.S. Department of Energy-National Energy Research Scientific Computing Center,
Particle astrophysics research computing center at the University of Maryland,
Institute for Cyber-Enabled Research at Michigan State University,
Astroparticle physics computational facility at Marquette University,
NVIDIA Corporation,
and Google Cloud Platform;
Belgium {\textendash} Funds for Scientific Research (FRS-FNRS and FWO),
FWO Odysseus and Big Science programmes,
and Belgian Federal Science Policy Office (Belspo);
Germany {\textendash} Bundesministerium f{\"u}r Forschung, Technologie und Raumfahrt (BMFTR),
Deutsche Forschungsgemeinschaft (DFG),
Helmholtz Alliance for Astroparticle Physics (HAP),
Initiative and Networking Fund of the Helmholtz Association,
Deutsches Elektronen Synchrotron (DESY),
and High Performance Computing cluster of the RWTH Aachen;
Sweden {\textendash} Swedish Research Council,
Swedish Polar Research Secretariat,
Swedish National Infrastructure for Computing (SNIC),
and Knut and Alice Wallenberg Foundation;
European Union {\textendash} EGI Advanced Computing for research;
Australia {\textendash} Australian Research Council;
Canada {\textendash} Natural Sciences and Engineering Research Council of Canada,
Calcul Qu{\'e}bec, Compute Ontario, Canada Foundation for Innovation, WestGrid, and Digital Research Alliance of Canada;
Denmark {\textendash} Villum Fonden, Carlsberg Foundation, and European Commission;
New Zealand {\textendash} Marsden Fund;
Japan {\textendash} Japan Society for Promotion of Science (JSPS)
and Institute for Global Prominent Research (IGPR) of Chiba University;
Korea {\textendash} National Research Foundation of Korea (NRF);
Switzerland {\textendash} Swiss National Science Foundation (SNSF).

\end{document}